\documentclass[%
 reprint,
 amsmath,amssymb,
 aps,
]{revtex4-2}

\usepackage{graphicx}
\usepackage{dcolumn}
\usepackage{bm}

\begin{document}

\preprint{APS/123-QED}

\title{Energy Gaps and Plateau Characteristics in the Fractional Quantum Hall Effect Derive from  Multi-particle Correlations}

\author{Jongbae Hong}
\email{jbhong@snu.ac.kr}
\affiliation{%
 School of Physics and Astronomy, Seoul National University, Seoul 08826 {\sc Korea}\\
 Asia Pacific Center for Theoretical Physics, Pohang, Gyeongbuk 37673 {\sc Korea}}%





\date{\today}

\begin{abstract}
The energy gaps appearing in the fractional quantum Hall effect (FQHE) remain an essential aspect of the investigation. Moreover, the plateau widths in the Hall resistance have been considered simply an effect of disorder as in the integral quantum Hall effect. The existing theories could neither explain the Hall resistance curve owing to plateau widths nor calculate the energy gaps. This study reveals that both the energy gaps and plateau widths contain fundamental many-body aspects of the FQHE. They are found to be connected via the strengths of multi-particle correlations, which do not affect the plateau heights. They are automatically quantized just by the presence of multi-particle correlations. This work focuses on correlated skipping electrons moving through the edge of an incompressible strip formed within a Hall bar. Consequently, a single-particle Hamiltonian was constructed incorporating the Zeeman energies of multiply-correlated skipping electrons. The resulting energy spectrum exhibits hierarchical splits of the Landau levels according to correlation order. The lowest Landau level is examined. Based on such level splitting, a previously measured Hall resistance curve and energy gaps are quantitatively explained by determining the parameters that describe the degrees of multi-particle correlations. The chemical potential and effective $g$-factors are additionally predicted for the Hall resistance. Furthermore, the fractional electron charge $e/(2n+1)$ for an electron participating in $n$-particle correlation was obtained by identifying the Fermi distribution function of $n$ correlated basic transport entities moving through the edge of the incompressible strip. Finally, the ideal-like Hall resistance was obtained at half-filling using the strengths of multi-particle correlations given in a regular pattern.
\end{abstract}


\maketitle


\section{\label{sec:level1}Introduction 
}

Two-dimensional electron systems often exhibit marvelous phenomena \cite{Klitzing, Tsui, Bednorz},
with fractional quantum Hall effect (FQHE) being one \cite{Tsui}.
In a clean two-dimensional multi-layer semiconductor system at low temperature under a strong perpendicular magnetic field, many plateaus can be observed at various fractional fillings in the Hall resistance.
Although certain plateaus occur at even-denominator fractions, for example, $5/2$ filling, at higher Landau levels \cite{Willet, Pan}, 
the lowest Landau level (LLL) has plateaus only at odd-denominator fractions.
In this study, we focus on the LLL.

Currently, it is believed that the energy gaps of a fractional quantum Hall system (FQHS) are also attributable to the many-body effect of interacting electrons as well as the many plateaus in the Hall resistance.
However, despite forty years of intense investigation into FQHE, the energy gap remains an ongoing issue~\cite{Shayegan,Pan2020}. 
Moreover, the existing theories have been unable to calculate the values of the energy gaps.

The Hall resistance, one of the fundamental quantities of the FQHE, must be considered in a reasonable manner based on multi-particle correlations, which are specific mechanisms presenting the many-body effect. 
The Hall resistance is characterized by plateau heights and widths.
The word "plateau" has been regarded as referring only to plateau height, while the plateau width has been neglected owing to it being considered an effect of disorder.
However, the plateau heights are quantized in a regular manner, $h/\nu e^2$, via a fractional number $\nu$ indicating the filling factor.
In other words, the plateau heights are too simple to carry complicated information reflecting the strengths of the multi-particle correlations. 
Thus, the plateau width may be the characteristic that can carry such information.
Therefore, this study considers the plateau width and energy gap as equally important fundamental quantities of the FQHE that reflect the strengths of multi-particle correlations. 

This study reveals that the plateau heights are automatically quantized into $h/\nu e^2$ just by the presence of the multi-particle correlations and that the plateau width and energy gap are connected via the strengths of the multi-particle correlations.
Consequently, a more in-depth understanding of the FQHE can be obtained from the same.
Therefore, clarifying the discovery of the abovementioned plateau characteristics and relationship between the plateau width and energy gap is the primary goal of this study, which was substantially supported by reproducing a measured Hall resistance and energy gaps.
Reproduction of the Hall resistance curve was done by predicting the chemical potential and effective g-factor of each plateau state of filling $\nu$.

\begin{figure*}[t]
\centering
\includegraphics{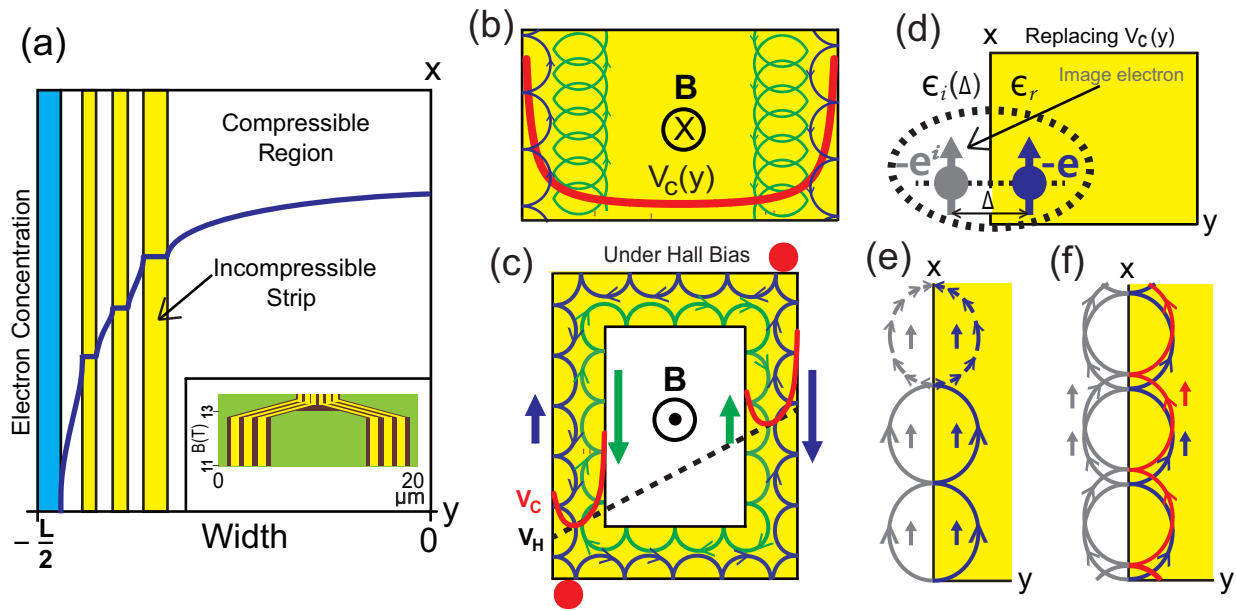}
\caption{(a) Electron concentrations in ISs (yellow), CSs (white), and a depletion region (blue) in the left half of a Hall bar of width $L$ are presented \cite{Lier}. 
The inset depicts a schematic of ISs in the region of a Hall current (brown) \cite{Gauss}.
(b) Semiclassical picture of electron motion in a FQHS: Skipping motion in a well-type confining potential $V_C(y)$ within an IS (blue) and a trajectory following a deformed cycloid away from the IS edge (green). 
(c) Left--right symmetric Hall currents (longer arrows) are formed by tilted confining potential in an IS loop due to a Hall bias $V_H$. Only the skipping trajectories are sketched. Hot spots
(red dots) were observed at the drain and source points \cite{Gauss}. 
(d) The confining potential $V_C(y)$ is replaced with a pair of an electron (blue) and its image (gray) of charge $-e^{\it i}$ and the same spin orientation. 
Electric permittivities $\epsilon_r$ and $\epsilon_i (\Delta)$ are given on the real and image electron sides, respectively, where $\Delta$ denotes the distance between the electron and its image.
(e) Semiclassical transport of a BTE through the edge of the IS. 
The varying rotational speed is illustrated in the upper part.
(f) Semiclassical transport of two correlated BTEs with four up spins.}
\end{figure*}

To accomplish the goal, this study attempts to construct an easily diagonalizable Hamiltonian containing the effect of multi-particle correlations among the electrons forming the Hall current.
However, the prevailing theories such as Laughlin wavefunction \cite{Laughlin} and composite fermion theory \cite{Jain}, which are employed to study pure two-dimensional electron systems, cannot calculate energy gaps~\cite{Shayegan,Du} and are unable to explain the plateau widths of a measured Hall resistance~\cite{Willet,Eisenstein} although the former predicts fractional electron charges that leads to fractional statistics in two dimensions~\cite{Halperin}, and the latter accommodates most of the plateau heights. 
Thus, the pure two-dimensional electron system may not be an appropriate model to fully explain the FQHE.
Therefore, accumulated studies reporting a realistic feature of a FQHS exhibiting an internal structure with an alternating arrangement of compressible and incompressible strips (ISs) due to background ions~\cite{Chklovskii, Lier, Lai, Ito, Suddards} were focused upon.
These two types of strips can be distinguished by their electron density profiles in the Hall bar, as shown in Fig.~1(a), which has been theoretically predicted~\cite{Chklovskii, Lier} and experimentally observed~\cite{Lai, Ito, Suddards} for an integral quantum Hall system (IQHS). 
According to the above theoretical studies, the IS exhibits an insulating nature owing to a constant electron density, while the compressible strip (CS) is conducting and provides a constant potential via charge redistribution.

Considering the realistic model comprising an alternating arrangement of ISs and CSs facilitates the study of FQHE from a different perspective and allows the construction of an easily diagonalizable Hamiltonian containing the effect of multi-particle correlations. 
In a two-dimensional electron system under a perpendicular magnetic field, the electron group velocity in the $x$-direction is proportional to the derivative of the potential in the $y$-direction. Thus, the Hall current flows only through the IS owing to the potential in the conducting CS being constant. Fast electrons follow a skipping trajectory at the IS edge, whereas relatively slow electrons follow a deformed cycloidal trajectory away from the IS edge, as shown in Fig.~1(b).
It is evident that the direction of drift is opposite at opposite side of the IS.
The consideration of the drifting speeds and directions depicted in Fig.~1(b) suggests the existence of a well-type confining potential $V_C(y)$ (red curve) within the IS.
The high potentials of $V_C(y)$ at both edges of the IS may originate from the accumulation of free electrons at the boundaries of the adjacent CSs, which are conducting. 

Recently, an important experiment~\cite{Gauss} that measured a left--right symmetric Hall current distribution over a width of about $5 \mu$m in a FQHS has been reported, as schematically shown in the inset of Fig.~1(a).
This wide region of Hall current can be understood by stacks of multiple ISs, as in the yellow strips in the same inset.
In contrast, the symmetric Hall current can be explained through the application of a Hall bias $V_H$, as shown in Fig.~1(c), wherein a Hall bias $V_H$ results in a tilted confining potential in the IS (red curve) and yields different group velocities (potential slopes) at opposite sides of the IS. The net current is given by the difference in group velocities (thick arrows), which occurs symmetrically on the left and right sides of the IS loop, thereby indicating a left--right symmetric Hall current.
Hot spots were also observed (red dots) through which the net current flows in and out.
The existence of hot spots is inevitable for electrons having different speeds along the IS edge to produce a steady current.

Ultimately, the observations of Ref.~\cite{Gauss} are compatible with the existence of the IS, CS, and well-type confining potential $V_C(y)$ within the IS, despite the calculations of 
Refs.~\cite{Chklovskii, Lier} not indicating the well-type confining potential explicitly. 
Thus, the study is continued based on the existence of the confining potential $V_C(y)$ within the IS.

\section{Effective Hamiltonian}
Constructing an effective Hamiltonian of the FQHS and obtaining its eigenvalues are the first tasks undertaken of this study.
The crucial energy, which is not considered in an ideal Hamiltonian, is the correlation energy that reflects the many-body effect among the electrons in the Hall current flowing through the IS edge.
The correlation energy is included in the Hamiltonian through the addition of the Zeeman energies of the skipping electrons with two-particle, three-particle correlations, and onwards.
For convenience, this study considers only the skipping motion representing the electron edge motions within the IS. 

To express the Zeeman energies induced by multi-particle correlations, first, the confining potential $V_C(y)$ in Fig.~1(b) is replaced with an image electron of charge $-e^{\it i}$ having the same spin orientation, as shown in Fig.~1(d).
An electric permittivity $\epsilon_i (\Delta)$ depending on the distance between the electron and its image is introduced on the image side.
Accordingly, the image charge is given by 
$-e^{\it i}(\Delta)=-e[\epsilon_r-\epsilon_i (\Delta)]/[\epsilon_r+\epsilon_i (\Delta)]$, where $\epsilon_r>\epsilon_i (\Delta)$ is the permittivity on the real electron side, following the image charge method of electrostatics~\cite{Jackson}.
Subsequently, the fitting of the Coulomb potential due to the image charge with $V_C(y)$ at the position of electron can be obtained by varying 
$\epsilon_i (\Delta)$.
The introduction of the image electron of the same spin orientation ensures electron wavefunction nodes at the boundaries of the IS via the Pauli principle, which bounce electrons back at the boundary.

Now, the problem turned into a form where an electron and its image follow mirror symmetric deformed semicircular trajectories, as depicted in Fig.~1(e), and a pair of skipping real and image electrons constitutes the basic transport entity (BTE) of the edge current within the IS, which may be considered as the quasiparticle of the IS edge state.
Further, correlated BTEs are considered, as illustrated in Fig.~1(f) having two-particle correlation, for instance. 
Such correlations are likely to form only among electrons drifting with equal group velocity, as correlation is strongly affected by inter-particle distance.
Therefore, considering $n$ correlated BTEs reflects the overall many-body effect in studying the FQHE.

As the magnetic energy of an individual BTE is expressed in the Zeeman energy form $-\vec{\mu}_{1}\cdot\vec{B}$, that of $n$ correlated BTEs is given by  $-\vec{\mu}_{n}\cdot\vec{B}$. 
Thus, the effective Hamiltonian in the single-particle description can be constructed via the addition of the Zeeman energies per real electron, $-\vec{\mu}_{n}\cdot\vec{B}/n$, where $n=1, 2,\cdots$, to the ideal part as follows: 
\begin{equation}
   H \!\! = \!\! \sum_{i=1}^{N_{is}} \!\! \left[ \frac{(\vec{p}_i+e\vec{A}_i)^2}{2m_c^*} - |g^*|\, \mu_{\scriptscriptstyle B} \! \frac{\vec{s}_i}{\hbar} \cdot \vec{B} \! - \sum_{n=1}^{N_{is}} \! \frac{\vec{\mu}_{n,i}\cdot\vec{B}}{n} \right], \label{eqn:hamil}
\end{equation}
where $N_{is}$ denotes the number of real electrons participating in the edge current of the IS, $m_c^*$ is the cyclotron effective mass, $\vec{p}$ and $\vec{A}$ are the electron momentum and external vector potential, respectively, $|g^*|$ is the effective $g$-factor, $\mu_{\scriptstyle B}=\hbar e/2m_0$ is the Bohr magneton with bare electron mass $m_0$, $\vec{s}_i$ is the electron spin operator, and $\vec{\mu}_{n,i}$ denotes $\vec{\mu}_{n}$ of the $i$-th electron. 

Double counting of the spin and orbital angular momenta comprising the magnetic moment $\vec{\mu}_{1}$ of the individual BTE in Eq. (\ref{eqn:hamil}) must be reexamined because the spin angular momentum is used in the second term and the magnetic moment $\vec{\mu}_{0}$ formed by the orbital angular momentum of uniform circular motion is contained in the first term.
Therefore, the corresponding double counting can be avoided by eliminating the spin contribution in $\vec{\mu}_{1}$ and subtracting $\vec{\mu}_{0}$ from 
$\vec{\mu}_{1}$, thus $\vec{\mu}_{1}$  in Eq. (\ref{eqn:hamil}) should read $(\vec{\mu}_{1}-\vec{\mu}_{0})$ excluding the spin. 

\section{Angular Momentum Quantization and Energy Spectrum}
To obtain the energy spectrum from the Hamiltonian, the expression $\vec{\mu}_n=\gamma_n\vec{J}_{n}$ utilizing the gyromagnetic ratio $\gamma_n$, which reflects the effect of the confining potential $V_C(y)$, is employed.
Thus, if  the eigenvalues of the $z$-component of the total angular momentum $\vec{J}_{n}^z/\hbar$, namely $m_{j_n}$, is determined for the system shown in Fig.~1(c), the Zeeman energy in the last term of Eq. (\ref{eqn:hamil}) is written as $-\vec{\mu}_{n}\cdot\vec{B}=-\hbar\gamma_n m_{j_n}B$ by setting $\vec{B}=B\hat{z}$, and the energy spectrum for the LLL becomes 
\begin{equation}
   E^\nu_{m_j}=\hbar\omega_c \left(\frac{1}{2} -\zeta_\nu-\sum_{n=1}^{N_{is}}\delta_{n}^\nu m_{j_n} \right), 
\label{eqn:eigenvalue}
\end{equation}
where $\omega_c=eB/m_c^*$, $\zeta_\nu=(|g^*_\nu|/4)(m_c^*/m_0)$, and 
$\delta_n^\nu=\hbar\gamma_n^\nu/2n\mu_{\scriptstyle B}^*$ with $\mu_B^*=\hbar e/2m^*_c$. 
Here, the parameter $\delta_n$ carries information on the strength of $n$-particle correlation through the gyromagnetic ratio $\gamma_n$.  
The filling factor $\nu$ is explicitly added in Eq. (\ref{eqn:eigenvalue}) because each plateau state of filling $\nu$ belongs to a different phase~\cite{Sondhi}.
Here a notable point is that determining the eigenvalues $m_{j_n}$ is the key step to explain the essential features of the FQHE through the energy spectrum. 

It is believed that the angular momentum in a system with a boundary is not equal to that in a free space. 
Therefore, care must be taken when a boundary exists in the system under consideration.
If an infinite square-well confining potential is considered in the IS loop in Fig.~1(c) with $V_H=0$, the role of boundary is just to move the rotating center along the boundary, which yields additional Zeeman energy. 
However, the additional magnetic energy produced by the counterclockwise motion (blue line) of the rotating center is cancelled by that formed by the clockwise motion (green line) at the opposite side of the strip if the two trajectories are close enough.
A narrow enough IS loop is assumed in this study.
Thus, the relevant angular momentum is that of the rotational part only. 
Since the angular momentum is defined at a point, the angular momentum at a point of skipping trajectory within the square-well potential is equal to that of the uniform circular motion in the LLL, whose angular momentum quantum number $\ell_1$ is unity~\cite{Tong}, which is clarified by the Laughlin wavefunction at the LLL for noninteracting electrons.

However, when the confining potential $V_C(y)$ exists, as shown in Fig.~1(b), the speed of rotation changes and the orbit is deformed, as illustrated in Fig.~1(e). 
The varying-speed rotational motion may be described by a linear combination of the eigenstates with magnetic quantum numbers $m_{\ell_1}=-1, 0, 1$, while the adiabatic deformation of the orbit due to $V_C(y)$ does not alter the angular momentum quantization, namely keeping $\ell_1=1$.
With such information, angular momentum addition was performed for the 
spin and orbital angular momenta comprising the total angular momentum $\vec{J}_{n}$ of $n$ correlated BTEs to subsequently obtain the magnetic quantum number $m_{j_n}=-n, \cdots, n$, namely the eigenvalues of the $z$-component of the total angular momentum $\vec{J}_{n}^z/\hbar$.
The procedure of angular momentum addition is presented in Appendix A.

Figure~2 shows the split Landau levels given in the last term of Eq. (\ref{eqn:eigenvalue}) depending on the correlation order $n$. 
As evident from Fig.~2, the odd-denominator filling states $\nu=p/(2n+1)$, where $p=1, \cdots, 2n$, arise from $n$-particle correlation; the size of the split gap governed by $\delta_n$ indicates the strength of $n$-particle correlation, and the half-filling state is derived from taking zero eigenvalues of $\vec{J}_n^z$, namely $m_{j_n}=0$, for all correlation orders, which yields zero correlation energy although the strengths of multi-particle correlations $\delta_n$ do not vanish at half-filling.

Based on the level splits shown in Fig.~2, the physics of FQHE revealing the role of multi-particle correlations in determining the energy gaps and plateau characteristics can be studied.

\section{Hall Resistivity Formula}
Hall resistance $R_H$ is given by the ratio of transverse Hall bias $V_y $ to longitudinal Hall current $I_x$. In a two-dimensional quantum Hall system with width $\Delta L$, which may be considered as the width of the IS, 
$R_H=V_y/I_x=E_y\Delta L/J_x\Delta L=\rho_{yx}$.
Therefore, the Hall resistance and the Hall resistivity are equivalent in the two-dimensional Hall effect.
Hereinafter, the term ``Hall resistivity'' is used.

The Hall resistivity expression $\rho_{xy}=B/e\rho_c=-\rho_{yx}$, where $\rho_c$ denotes the carrier density, can be easily derived using the definition of current density $\vec{J}=-e\rho_c\vec{v}
=\rho^{-1}\vec{E}$ and the equation of motion with a Drude term at a steady-state $-e\vec{E}-e\vec{v}\times\vec{B}-m_0\vec{v}/\tau=0$, where $\tau$ denotes the scattering time, expressed in a matrix form with $\vec{v}=(v_x, v_y, 0)$, $\vec{E}=(E_x, E_y, 0)$, and $\vec{B}=(0, 0, B_z)$~\cite{Tong}.
Therefore, obtaining the quantum Hall resistivity is equivalent to calculating the field-dependent carrier density $\rho_c(B)$, which is given by 
$\rho_c(B)=D^L\sum_{\{m_j\}}f_{m_j}$, where $D^L = eB/h$ is the unit-area Landau level degeneracy and $f_{m_j}$ is the Fermi distribution function comprising the energy spectrum $E_{m_j}(B)$ of Eq. (\ref{eqn:eigenvalue}) and field-dependent chemical potential.

\begin{figure}[t]
\centering
\includegraphics[width=1.0\linewidth]{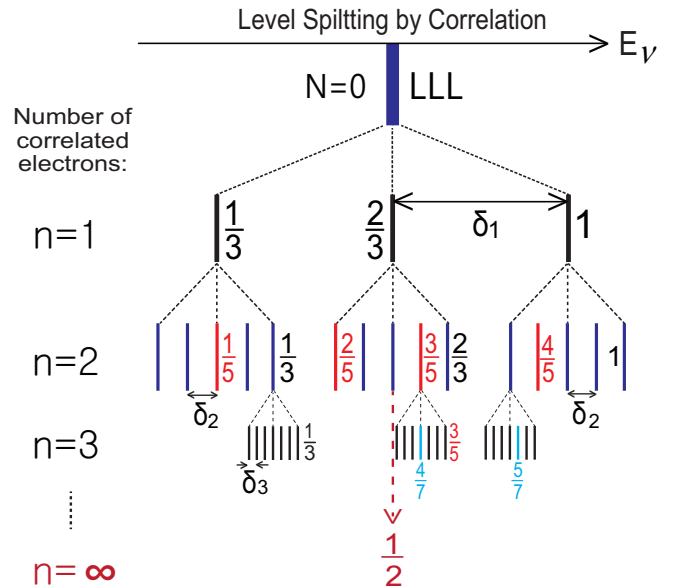}
\caption{Schematic of LLL splitting by the additional Zeeman interactions in Eq. (\ref{eqn:eigenvalue}). 
The correlation order $n$ is shown on the left side. 
The $\delta_n$ terms denote the split by $n$-th order correlation. The half-filling state is reached at the limit of the splitting hierarchy.}
\end{figure}  

The curly brackets in the summation $\sum_{\{m_j\}}$ for energy eigenvalues were used to denote the multiplicity of the eigenvalue $m_{j_n}$.
The level splits shown in Fig.~2, which are driven by the multiplicity of $m_{j_n}$, require to write the summation over the Fermi distribution function as follows:
\begin{equation}
\sum_{\{m_{j}\}}f_{m_j} = \frac{1}{3} \sum_{m_{j_1}=-1}^{+1}f_{m_{j_1}}\times \frac{1}{5}\sum_{m_{j_2}=-2}^{+2}f_{m_{j_2}}\times\cdots, 
\label{eqn:fermi}
\end{equation}
where each dividing number corresponds to the multiplicity of $m_{j_n}$.
Dividing the multiplicity amounts to rendering the upper bound of the sum over Fermi distribution function, $\sum_{\{m_{j}\}}f_{m_j}$, unity, which is explicitly shown in Fig.~3(a).
Thus, the field-dependent Hall resistivity $\rho_{xy}(B)$ is written as
\begin{equation}
 \rho_{xy}(B)=(h/e^2) \Big/ \sum_{\{m_j\}}[1+{\rm exp}\{(E_{m_j}-\mu)/k_BT\}]^{-1}.
\label{eqn:resistivity}
\end{equation}
An explicit expression of the denominator considering up to two-particle correlation is presented in Appendix B.
Toyoda \cite{Toyoda} used this form to obtain the field-dependent Hall resistivity using an ad hoc energy spectrum. 

Equation (\ref{eqn:fermi}) is composed of multiple components corresponding to the BTEs with different numbers of correlated electrons.
Therefore, it is appropriate to consider the component $\frac{1}{2n+1}f_{m_{j_n}}$ as a Fermi distribution function for electrons participating in $n$ correlated BTEs.
This point of view for Eq. (\ref{eqn:fermi}) is used in discussing the fractional electron charges in Sec. VI.

\section{Results}
This section qualitatively shows how the plateau characteristics, namely the plateau width and height, are determined and quantitatively explains the energy gaps~\cite{Du} and Hall resistivity~\cite{Eisenstein}. 
However, obtaining the Hall resistivity using the formula presented in Eq. (\ref{eqn:resistivity}) involves the determination of the chemical potential.
Therefore, before calculating the Hall resistivity, the chemical potential is determined first.

\subsection{Plateau Characteristics}
The plateau characteristics were clarified by performing a simple trial calculation of field-dependent carrier density $\rho_c(B)$ considering up to two-particle correlation ($n=1$ and 2), which contains fifteen Fermi distribution functions, as shown Appendix B.

\begin{figure}[t]
\centering
\includegraphics[width=0.9\linewidth]{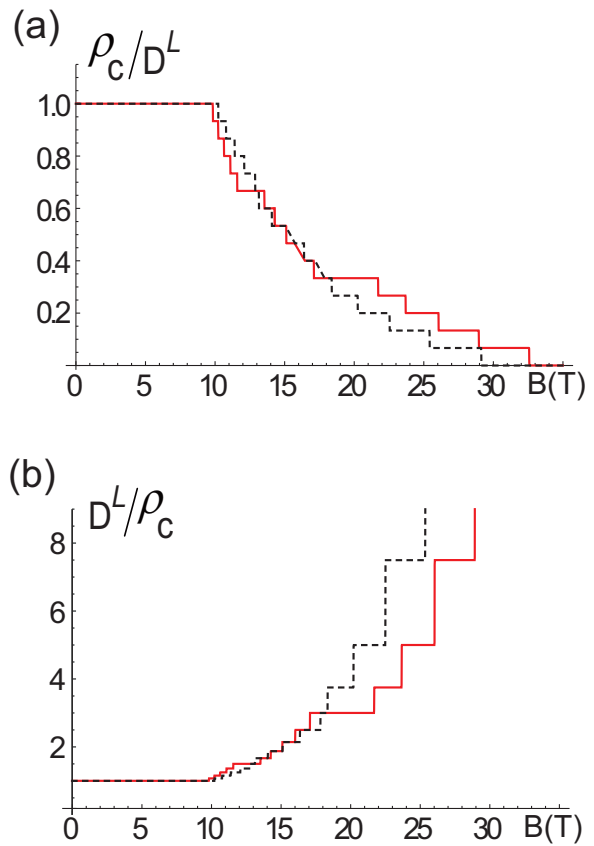}
\caption{(a) The carrier density of fifteen equal-height stairs:  $\delta_1=0.21$ and $\delta_2=0.029$ (red); $\delta_1=0.164$ and $\delta_2=0.038$ (dashed). $\mu=13$ meV, $m_c^*/m_0=0.067$, $g*=0$, and $T=10$ mK were set.
(b) Inverse of the lines in (a). }
\end{figure}

Two field-dependent carrier densities discriminated by different $\delta_n$ were compared in Fig.~3(a). Both are composed of fifteen equal-height stairs indicating fifteen Fermi distribution functions. 
It is evident that the stair widths depend on the values of $\delta_n$, whereas the stair heights of $1/15$ are robust. 
Thus, the vertical positions of the flat parts exactly coincide with the filling fraction $\nu$.
The Hall resistivity $\rho_{xy}(B)$, expressed in units of $h/e^2$, is the very $D^L/\rho_c(B)$ given in Fig.~3(b), wherein each plateau at filling $\nu=1, \frac{14}{15},\cdots,\frac{2}{15}$, and $\frac{1}{15}$ from low magnetic field amounts to the inverse of the corresponding flat region in Fig.~3(a).
Further, the denominator 15 amounts to the number of split levels shown in the row of $n=2$ in Fig.~2, which is equal to the total cross-multiplicity produced by the single ($n=1$) and two correlated BTEs ($n=2$).

Consequently, the plateau characteristics are identified as follows: Plateau widths are determined based on the strengths of multi-particle correlations $\delta_n$, while plateau heights, although independent of the strengths, are automatically fixed at Hall resistivity $(h/e^2)/\nu$, where $\nu=\theta/(2n+1)!!$ with $\theta=1,2, \cdots,(2n+1)!!$, once multi-particle correlation includes up to $n$ particles.
These qualitative statements regarding the plateau characteristics can be substantially supported by the quantitative reproduction of a measured Hall resistivity curve and energy gaps. 

\begin{figure}[t]
\centering
\includegraphics[width=1.0\linewidth]{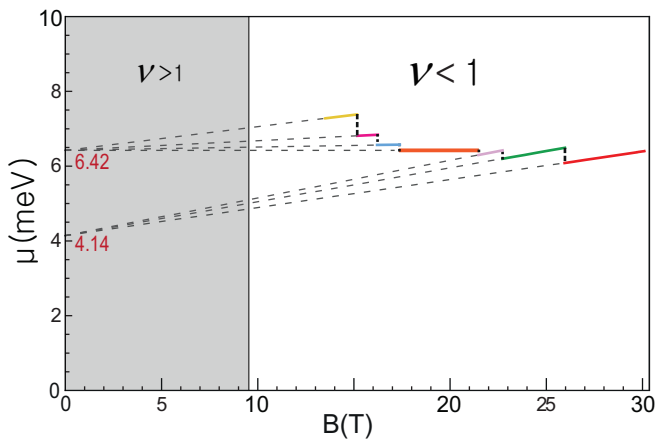}
\caption{Piece-wise linear chemical potential with $\mu_{1/2}=6.42$ meV and $\mu_{1/4}=4.14$ meV. Each zone is matched to the Hall resistivity interval in Fig.~6(a) according to color. } 
\end{figure}

\subsection{Chemical potential} 
The chemical potential was set to increase linearly and drop at a transition point over the range of magnetic fields in each plateau state. Such behavior has been observed experimentally in both an IQHS \cite{Wei} and FQHS \cite{Kharapai}.
Therefore, the field-dependent chemical potential $\mu(B)$ was set as a piece-wise linear function of the external magnetic field of the form $\mu_\nu (B) =\kappa_\nu {B}+\mu^0_\nu$ between the transition points of a plateau state of filling $\nu$, where $\kappa_\nu$ and $\mu^0_\nu$ denote the slope and intercept at $B=0$ of the chemical potential, respectively. 
Further, $\kappa_\nu$ was determined from the extent of the drop, which corresponds to the energy gap, as mentioned in Ref.~\cite{Kharapai2}. Here, the energy gap data provided in Ref.~\cite{Du} was employed to determine $\mu_\nu(B)$, as shown in Fig.~4, by choosing the reference $\mu^0_\nu$ as the chemical potential at an appropriate even-denominator filling. 
In addition, for the sequence $\nu = q/(2q-1)$ with $q=2, 3,\cdots$ that corresponds to the upper part of Fig.~4, $\mu_\nu^0=\mu_{1/2}$ was chosen. 
However, a different $\mu_\nu^0$ is needed for the lower part of Fig.~4 that belongs to the sequence $\nu = q/(2q+1)$ with $q=1, 2,\cdots$.
Here, $\mu^0_{\nu}=\mu_{1/4}$ was chosen to form a reasonable chemical potential structure.
$\mu_{1/2}$ and $\mu_{1/4}$ were determined using the energy spectrum of Eq.~(\ref{eqn:eigenvalue}) and obtaining the ideal-like Hall resistivity for these even-denominator fillings. 
Details regarding the same are provided in Appendix C.
The values of $\kappa_\nu$ in units of meV/T and $\mu_\nu^0$ in units of meV are listed in Table I.

The effective $g$-factor contained in $\zeta_\nu$ in Eq.~(\ref{eqn:eigenvalue}) must be determined along with the chemical potential because both $\mu_\nu(B)$ and $\hbar\omega_c(B)\zeta_\nu$ are linear in $B$, and they only shift the Hall resistivity horizontally.
Because $\mu_\nu(B)$ is nearly determined using the experimental data for energy gap as mentioned previously, only fine tuning was needed to fix it, as shown in Fig.~4. 
Therefore, determining $\zeta_\nu$ is comparatively easy.

\subsection{Hall resistivity and Energy gaps}
The Hall resistivity is obtained using the formula given in Eq. (\ref{eqn:resistivity}). 
In contrast, the definition of the energy gap in the FQHE is still unclear.
In this study, it was defined as the transition energy from the filled state of filling $\nu$ to the lowest empty state, as suggested by Rey through advisory discussion.
Thus, the energy gap can be expressed as $E^\nu_G = \xi \Delta_\nu B_\nu$, where $\xi=e\hbar/m_c^*$, and $\Delta_\nu$ is a dimensionless gap parameter obtained from a combination of $\delta_n^\nu$, as shown in Table I. 
$E^{\frac{1}{3}}_G$ is shown in Fig.~5 using Fig.~2.

Now, it is clear that the energy gap is connected to the plateau width via the strengths of multi-particle correlations. 
The $\delta_n$-dependence of plateau width was explicitly shown in Fig.~3.
Therefore, the most crucial parameters in obtaining the Hall resistivity and energy gaps are $\delta_n^\nu$, which were naturally fixed by fitting the two independent experimental results~\cite{Eisenstein, Du} simultaneously for a given filling $\nu$, thus restricting the freedom of $\delta_n^\nu$.
In this study, $m_c^*/m_0 = 0.059$ was set, which is close to the value of GaAs. Consequently, $\xi=1.9613$ meV/T was obtained.

\begin{figure}[t]
\centering
\includegraphics[width=1.0\linewidth]{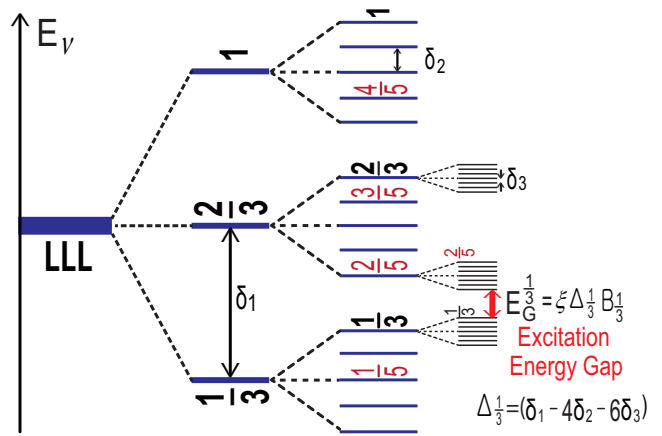}
\caption{ Illustration of the energy gap $E_G$ at filling $\nu =\frac{1}{3}$ from Fig.~2.
} \label{fig:Fig5}
\end{figure}

\begin{figure*}[t]
\centering
\includegraphics[width=.9\linewidth]{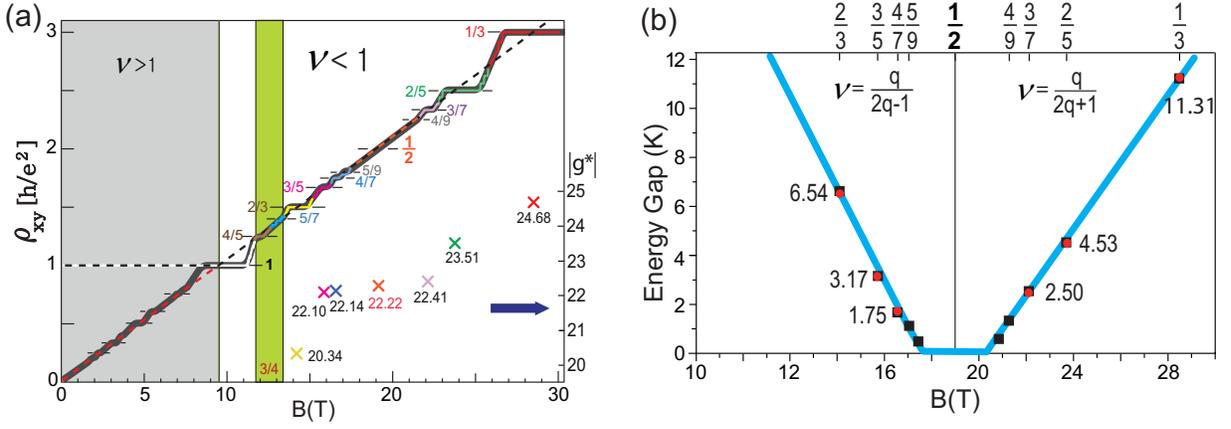}
\caption{(a) Theoretical Hall resistivity superimposed on the experimental data (thick gray line) \cite{Eisenstein} along with effective $g$-factors (cross marks).
The black dashed line is the theoretically obtained Hall resistivity at $\nu=\frac{1}{2}$ and the red dashed line is its extrapolation. The green region belongs to the filling sequences $\nu=(3q\pm 1)/(4q\pm 1)$ that include $\nu=\frac{4}{5}$ and $\nu=\frac{5}{7}$. 
(b) Theoretical energy gaps (red dots) are superimposed on the experimental data (black squares) reported in Ref. \cite{Du}.
The blue line is our prediction over the entire magnetic field range.
} 
\label{fig:Fig6}
\end{figure*}

\begin{table*} [ht]
\centering
\caption{Values of $\kappa_\nu$, $\mu_\nu^0$, $\zeta_\nu$, $\delta_n^\nu$, and the magnetic field along with energy gap parameter $\Delta_\nu$. $\lambda$ = 34.46. }
   \vspace{0.1cm}
    \setlength{\tabcolsep}{5pt}
    \begin{tabular}{c|cccccccccc}
    \hline\hline
\\[-2ex]
 {\small $\nu$} & {\small 1/3} & {\small  2/5} & {\small 3/7} & {\small 1/2} & {\small 4/7}
& {\small 3/5} & {\small 2/3} & {\small 5/7} & {\small 4/5} & {\small 1} \\[0.3ex]
 Color & Red & Green  &  Violet &  Orange   & Blue & Magenta  & Yellow  & Cyan   & Brown  & White \\[0.3ex]
\cline{1-11}
\\[-2ex]
  $10^2\cdot\kappa_\nu$ & 7.495 & 9.059 & 10.049 & 0 & 0.9075 & 2.612 & 6.352 & 7.800 & 3.189 & 9.500 \\[0.3ex] \hline \\[-2ex]
  $\mu_\nu^0$ & 4.14 & 4.14 & 4.14 & 6.42 & 6.42 & 6.42 & 6.42 & 6.42 & 7.99 & 9.30 \\[0.3ex] \hline \\[-2ex]
  $10^2\cdot\zeta_\nu$ & 36.41 & 34.68 & 33.05 & 32.77 & 32.65 & 32.60 & 30.00 & 30.60 & 27.20 &  7.70 \\[0.3ex] \hline \\[-2ex]
  $10^2\cdot\delta_1^\nu$ & 5.410  & 5.830 & 9.050 & $\lambda$/3         & 11.73 & 12.06 & 12.30 & 13.40  & 14.30 & 3.250  \\[0.3ex]
  $10^2\cdot\delta_2^\nu$ & 0.810  & 1.295 & 2.137 & $\lambda$/15       & 2.540 & 3.100 & 2.400 & 2.400  & 3.000 & 0.6535  \\[0.3ex]
  $10^2\cdot\delta_3^\nu$ & 0.071   & 0.076  & 0.600  & $\lambda$/105 & 0.657  & 0.350 & 0.110 & 0.560  & 0.300 & 0.0756 \\[0.3ex]
  $10^2\cdot\delta_4^\nu$ & 0.000  & 0.000   & 0.013 & $\lambda$/945  & 0.024 &  0.014 &  0.000 &0.000  & 0.000 & 0.000 \\[0.5ex]
\hline\hline\\[-2.0ex]
    & \ \ $\delta_1$    &        0      &        0    &  0 &        0     &       0      &    \ \  $\delta_1$  &  $-$  & $-$  &  $-$ \\[0.5ex]
    & $-4\delta_2$  & \ \ $\delta_2$   & 0  & 0 & 0  &$\ \ \delta_2 $ & $-4 \delta_2$ &$-$ &$-$ & $-$ \\[0.5ex]
  $\Delta_\nu$       & $-6\delta_3$  & $-6\delta_3$ &$ \ \ \delta_3$ & 0  &      $\ \ \delta_3$    &$-6\delta_3$ & $-6\delta_3$ &  $-$  &$-$ & $-$ \\[0.5ex]
   & $-8\delta_4$  & $-8\delta_4$ &$ - 8 \delta_4$ & 0  &     $- 8 \delta_4$    &$-8\delta_4$ & $-8\delta_4$ &$-$  &$-$ & $ - $ \\[0.5ex]
\hline \\[-2.0ex]
  $B({\rm T})$ & 28.50 & 23.75       & 22.12      & 19.00 & 16.57  & 15.71   & 14.10  &  13.30  &  12.00   & 9.50 \\[0.5ex]
\hline
\end{tabular}
\end{table*}

The Hall resistivity and energy gaps were calculated using the values listed in Table I and setting $T=10$ mK in calculating the Hall resistivity. Subsequently, the calculated results superimposed on previous experimental data \cite{Eisenstein,Du} are shown as Fig.~6(a) and 6(b), respectively. 
In the former plot, monotonically increasing effective $g$-factors with $B$ were added. 
Theoretical results of this work agree remarkably well with the experimental data over the entire range of magnetic field.
In addition, Table I indicates that only a few $\delta_n$ for each filling are sufficient to explain the Hall resistivity curve and energy gaps, although all possible correlations were considered in Eqs. (\ref{eqn:hamil}) and (\ref{eqn:eigenvalue}).
It is noteworthy that only half-filling has a regular pattern of $\delta_n$, $\delta_n=0.3446/[3\cdot 5\cdots(2n+1)]$, with which the ideal-like Hall resistivity [dark dashed line in Fig.~6(a)] was obtained. 
The specific procedure to obtain the Hall resistivity curve of Fig.~6(a) is shown in Fig.~7 in Appendix D. 

\section{Fractional Charges}
In this section, the fractional charges of the electron forming the edge current in the IS are discussed via identifying the Fermi distribution functions of individual and correlated BTEs.
As mentioned about the constituents of Eq. (\ref{eqn:fermi}) in the last part of Sec. IV, the summation over the Fermi distribution function $\sum_{\{m_j\}}f_{m_j}$ appearing in the Hall resistivity formula of Eq. (\ref{eqn:resistivity}) comprises the Fermi distribution functions $f_{m_{j_1}}/3$, $f_{m_{j_2}}/5$, $\cdots$, and $f_{m_{j_n}}/(2n+1)$, which respectively apply to electrons participating in an individual BTE, two correlated BTEs, $\cdots$, and $n$ correlated BTEs. 
This implies that when the system sets at one of fillings $\nu=p/5$ where $p=1, \cdots, 4$, for example, two correlated BTEs play a role, and the statistical average of the total charge $\langle Q\rangle$ contributing to the Hall current by $N$ electrons that are involved in two correlated BTEs is given by $\langle Q\rangle=\langle N/5 \rangle e$, which amounts to an electron having an effective charge of $e/5$.
In general, when the system sets at one of odd-denominator fillings $\nu=p/(2n+1)$, where $p=1, \cdots, 2n$, $n$ correlated BTEs play a role, and electrons have effective charge $e/(2n+1)$.
Under this circumstance, the leading splits in Fig.~2 must be $(2n+1)$ to satisfy the insulating condition of the IS.
Therefore, it is concluded that the effective electron charge is determined by the fractional  filling at which the system is set.
Fractional charges $e/3$ and $e/5$ have been observed \cite{Picciotto, Reznikov, Saminadayar}.

\section{Half-filling state}

The half-filling state is special, as shown in Fig.~2, in which the underlying LLL splits by cross-correlations among the BTEs with different numbers of correlated electrons.
The split levels in Fig.~2 equally share the LLL states of degeneracy $D^L =\frac{eB}{h}$.
Thus, each split level produced by considering up to $n$-body correlation has the level degeneracy $D_n^L=\frac{e}{(2n+1)!!}\frac{B}{h}$, which amounts to
having an effective charge $e^*_{n}=e/(2n+1)!!$ at the $n$-th level of correlation hierarchy shown in Fig.~2.
Although this effective charge is not a quasiparticle effective charge discussed in the last section, this hypothetical model is used in analyzing the half-filling state below.

It is meaningful to show that the ideal-like Hall resistivity (orange line) around half-filling in Fig.~6(a) is also an aspect of multi-particle correlation. 
The dark dashed line in Fig.~6(a) was obtained using the parameter values for $\nu=1/2$ in Table I, and it is well agreed with the orange part of the Hall resistivity around half-filling.
A notable fact is that the values of $\delta_n$ at half-filling were determined regularly, unlike other fillings, as shown in Table I.
The unique features of half-filling state can be extracted from the regular pattern of $\delta_n$. 
To extract them, a different expression of $\delta_n$ is used, which is obtained by putting the expression of gyromagnetic ratio, $\gamma_n={\tilde g}_n{\tilde \mu}_{nB}/\hbar$ with ${\tilde \mu}_{nB}=\hbar {\tilde e}^*_n/2{\tilde m}^*_n$, where ${\tilde g}_n$, ${\tilde\mu}_{nB}$, ${\tilde e}_n^*$, and ${\tilde m}_n^*$ denote the $g$-factor, Bohr magneton, effective charge, and effective mass of $n$ correlated BTEs, respectively, into the previous expression $\delta_n=\hbar\gamma_n/2n\mu_{\scriptstyle B}^*$ defined in Eq. (2).
The new expression of $\delta_n$ is written as
\begin{equation}
\delta_n=(1/2)g_n(e^*_{n}/e)(m_c^*/m_n^*), 
\end{equation} 
where $g_n={\tilde g}_n/n$, $e_n^*={\tilde e}_n^*/n$, and $m_n^*={\tilde m}_n^*/n$, indicating the quantities per electron involving in $n$ correlated BTEs.

From Kohn's theorem \cite{Kohn} asserting that the cyclotron effective mass is unaffected by interactions, the equality $m^*_n=m^*_c$ is taken, and the abovementioned effective charge $e^*_{n}=e/(2n+1)!!$ is applied to Eq. (5).
Then, the $\delta_n$ in Eq. (5) is given as $\delta_n=(g_n/2)/[3\cdot 5\cdots(2n+1)]$, which means that the effect of multi-particle correlation is solely contained in the $g$-factor $g_n$.
Comparing this expression with $\delta_n=0.3446/[3\cdot 5\cdots(2n+1)]$ (Table I) used in calculating the Hall resistivity at 
half-filling indicates that $g_n/2=0.3446$, which signifies that at half-filling, all orders of multi-particle correlation apply to the same degree of influence on the system.
This situation may create the same phenomenon produced by an ideal system without a correlation between the particles.
Thus, the ideal-like Hall resistivity was obtained even though the multi-particle correlation did not disappear.

Another ideal-like aspect of the half-filling state is the vanishing of the correlation energy, which was already mentioned at the last part of Sec. III.
The flat lines in the chemical potential (Fig.~4) and energy gaps [Fig.~6(b)] and the straight orange line in the Hall resistivity [Fig.~6(a)] respectively imply that the chemical potential drops, energy gaps, plateau widths near half-filling  are negligibly small. 

\section{Discussion}
This study was restricted to filling sequences $\nu=q/(2q\pm 1)$; extension to other fillings such as $\nu =q/(4q\pm 1)$, $\nu=(3q\pm 1)/(4q\pm 1)$, and $\nu=1$ could not be implemented owing to inadequate consistent experimental data.
However, the Hall resistivities for $\nu=\frac{5}{7}, \frac{4}{5}$, and $1$ were added for a continuation of the curve although their parameter values presented in Table I were less reliable than those of fillings $\nu=q/(2q\pm 1)$.

Another meaningful issue to discuss is the particle-hole asymmetry in the energy gaps shown in Fig.~6(b) as well as the recent data provided in Ref.~\cite{Shayegan} and its supplementary materials.
The origin of the particle-hole asymmetry was revealed to be the asymmetric distribution of the energy levels of the plateau states from the reference state $\nu=1/2$, as shown in Fig.~2.

\section{Conclusion}
This study constructed a nontrivial single-particle Hamiltonian via the addition of the Zeeman energies of the correlated skipping electrons that were moving through the edge of an incompressible strip in an FQHS based on experimental observations~\cite{Lai, Ito, Suddards, Gauss}.
Consequently, the energy spectrum was obtained via total angular momentum quantization of the correlated skipping electrons.
Both the plateau width and energy gap were considered as being equally important fundamental quantities that reflected the degrees of multi-particle correlations. In contrast, the plateau heights were exactly obtained simply by the presence of multi-particle correlations.
Moreover, the remarkable reproduction of an experimental Hall resistivity \cite{Eisenstein} and energy gaps \cite{Du} strongly supported the validity of the proposed Hamiltonian for the FQHS and coupling of the plateau width to the energy gap.
However, the correlation parameters $\delta_n^\nu$, slopes of chemical potential, and effective $g$-factors were determined phenomenologically to fit the experiments quantitatively.  
Furthermore, this study was restricted to the LLL, and extension to higher Landau levels, in which interesting plateaus occur at certain even-denominator fractions, remains an open challenge.

\begin{acknowledgments}
The author thanks A. Gauss, K. Kim, T. Toyoda, J. Weis, and S. H. Yoon for their help and valuable discussions, and particularly S.-J. Rey for sharing ideas on the energy gap and valuable comments on the manuscript.
This work was supported by a Korea National Research Foundation grant 2021R1I1A1A01040722 
and partially by the Open KIAS Center, KIAS, Korea.
\end{acknowledgments}

\appendix
\section{Total angular momentum quantization}
To make the energy spectrum available, the quantization of total angular momentum is needed. 
An appropriate scheme for angular momentum addition in this case is $jj$-coupling: first, summing the orbital and spin angular momenta for the electron and its image separately and then summing each total angular momentum. 
The original BTE including spins shown in Fig.~1(e) instead of the spinless BTE appearing in the $n=1$ term of Eq. (\ref{eqn:hamil}) is used to obtain the total angular momentum quantization for $n$ correlated BTEs.

The total angular momentum $\vec{J}_1$ of a single BTE is given by
$\vec{J}_1=(\vec{L}_1^{\,e}+\vec{s}_1^{\,e})+(\vec{L}_1^{\,i}+\vec{s}_1^{\,i})$, where 
$\vec{L}_1^{\,e} (\vec{L}_1^{\,i})$ and $\vec{s}_1^{\,e} (\vec{s}_1^{\,i})$ denote the orbital and spin angular momenta of the skipping electron (image), respectively.
It was mentioned in the text that the single BTE following a deformed semicircle with varying rotational speed is described by a linear combination of the eigenstates with respect to the eigenvalues of $(\vec{L}_1^{\,e,i})^z/\hbar$, i.e., $1, 0$, and $-1$.
In contrast, the allowed eigenvalue of $(\vec{s}_1^{\,e,i})^z/\hbar$ is only $\frac{1}{2}$ because the spins in the LLL are fully polarized~\cite{Kukushkin}. 
Therefore, the eigenvalues allowed in the addition of $(\vec{L}_1^{\,e,i})^z$ and $(\vec{s}_1^{\,e,i})^z$ are $m_{j_1}^{e,i}=-1+\frac{1}{2}, 0+\frac{1}{2}, 1+\frac{1}{2}$ for $(\vec{J}_1^{\,e,i})^z/\hbar$.
Now, adding $m_{j_1}^{e}$ and $m_{j_1}^{i}$ in all possible manner produces $-1, 0, 1, 2,$ and $3$, among which only the values satisfying the inversion symmetry of the eigenvalues of total angular momentum $z$-component are taken for the eigenvalues of $\vec{J}_1^z/\hbar$.
They are $m_{j_1}=-1, 0, 1$.
Consequently, the eigenvalues $m_{j_n}=-n, \cdots, n$ are obtained for $\vec{J}_n^z/\hbar$ of $n$ correlated BTEs via the same manner.

\section{Hall resistivity calculation }
The explicit expression of Hall resistivity $\rho_{xy}(B)=(B/e)\rho_c^{-1}(B)$ considering up to two-particle correlation, namely $n=1$ and $2$, is given by the expression of carrier density  $\rho_c(B)$, which is written as
\begin{equation*}
\begin{aligned}
&\rho_c(B)=\frac{eB}{h}\times \\
&\frac{3\times 5} {\sum_{m_{j_1}=-1}^{+1}\sum_{m_{j_2}=-2}^{+2}[1+e^{\beta
\{\hbar\omega_c(\frac{1}{2}-\zeta-m_{j_1}\delta_1-m_{j_2}\delta_2)-\mu\}}]^{-1}}, 
\end{aligned}
\end{equation*} 
where $\beta=1/k_BT$, using the energy spectrum $E_{m_j}$ given in Eq.~(\ref{eqn:eigenvalue}).
The explicit expression of the denominator is written as follows:
\begin{equation*}
\begin{aligned}
&(1+\alpha e^{-1*\delta_1\tilde{\gamma}}e^{-2*\delta_2\tilde{\gamma}})^{-1}
+(1+\alpha e^{-1*\delta_1\tilde{\gamma}}e^{-1*\delta_2\tilde{\gamma}})^{-1}\\
&+(1+\alpha e^{-1*\delta_1\tilde{\gamma}}e^{0*\delta_2\tilde{\gamma}})^{-1}
+(1+\alpha e^{-1*\delta_1\tilde{\gamma}}e^{1*\delta_2\tilde{\gamma}})^{-1}\\
&+(1+\alpha e^{-1*\delta_1\tilde{\gamma}}e^{2*\delta_2\tilde{\gamma}})^{-1}
+(1+\alpha e^{0*\delta_1\tilde{\gamma}}e^{-2*\delta_2\tilde{\gamma}})^{-1}\\
&+(1+\alpha e^{0*\delta_1\tilde{\gamma}}e^{-1*\delta_2\tilde{\gamma}})^{-1}
+(1+\alpha e^{0*\delta_1\tilde{\gamma}}e^{0*\delta_2\tilde{\gamma}})^{-1}\\
&+(1+\alpha e^{0*\delta_1\tilde{\gamma}}e^{1*\delta_2\tilde{\gamma}})^{-1}
+(1+\alpha e^{0*\delta_1\tilde{\gamma}}e^{2*\delta_2\tilde{\gamma}})^{-1}\\
&+(1+\alpha e^{1*\delta_1\tilde{\gamma}}e^{-2*\delta_2\tilde{\gamma}})^{-1}
+(1+\alpha e^{1*\delta_1\tilde{\gamma}}e^{-1*\delta_2\tilde{\gamma}})^{-1}\\
&+(1+\alpha e^{1*\delta_1\tilde{\gamma}}e^{0*\delta_2\tilde{\gamma}})^{-1}
+(1+\alpha e^{1*\delta_1\tilde{\gamma}}e^{1*\delta_2\tilde{\gamma}})^{-1}\\
&+(1+\alpha e^{1*\delta_1\tilde{\gamma}}e^{2*\delta_2\tilde{\gamma}})^{-1},
\end{aligned}
\end{equation*} 
where $\alpha=e^{\beta\{\hbar\omega_c(\frac{1}{2}-\zeta)-\mu\}}$ 
and $\tilde{\gamma}=\hbar\omega_c/k_BT$. 
The superscript $\nu$ in $\zeta$ and $\delta$ was deleted for convenience.
If correlation is considered up to four particles, the denominator has $945$ terms.

\section{Further information on chemical potential}
The chemical potential at half filling is given by  $\mu_{1 / 2}=\xi(\frac{1}{2}-\zeta_{1/2})B_{1/2}-E^{1/2}_{corr}$, where $\xi=e\hbar/m_c^*$, $\zeta_\nu=(|g^*_\nu|/4)(m_c^*/m_0)$, and $E^{1/2}_{corr}=\hbar\omega_c\sum_{n=1}^\infty\delta_{n}^{1/2} m_{j_n}$ from Eq.~(3) in the text. 
It is noteworthy that the correlation energy $E^{1/2}_{corr}$ vanishes at half filling even though $\delta_{n}^{1/2}\neq 0$. 
This is because the half-filling state is achieved by taking zero eigenvalues of $\vec{J}_n^z$, i.e., $m_{j_n}=0$, for all correlation orders $n$.
This situation was explicitly depicted in Fig.~2 of the text. Therefore, the chemical potential at half filling is given by $\mu_{1 / 2}=\xi(\frac{1}{2}-\zeta_{1/2})B_{1/2}$.
It is rather easy to fix $\zeta_{1/2}$ and $\mu_{1/2}$ using the condition that the ideal-like Hall resistivity must be recovered at half filling.    
The ideal-like Hall resistivity was obtained in terms of $\mu_{1 / 2}=6.420$ meV, $\zeta_{1/2}=0.32772$, and $\delta_n^{1/2}=0.3446/[3\cdot 5\cdots(2n+1)]$ (see Table I) at $B_{1/2}=19.0$ T.

The chemical potential at quarter filling, $\mu_{1 / 4}$, naturally contains the effect of the correlation energy. One can see that the location of quarter filling in Fig.~2 is away from half filling due to the correlation energy. Hence, $\mu_{1 / 4}$ is written as 
$\mu_{1 / 4}=\xi(\frac{1}{2}-\zeta_{1/4})B_{1/4}-E^{1/4}_{corr}$, where $E^{1/4}_{corr}=\xi(\delta_{1}-\delta_{2}-\delta_{3}+\cdots)B_{1/4}$ from Fig.~2. 
The ideal-like Hall resistivity using $\mu_{1 / 4}=4.140$ meV, $\zeta_{1/4}=0.3890$, and $\delta_n^{1/4}=0.222/[3\cdot 5\cdots(2n+1)]$ was recovered. 
However, an approximate calculation for $\mu_{1 / 4}$ was done by considering up to $\delta_3$ from $\delta_n^{1/4}=0.222/[3\cdot 5\cdots(2n+1)]$ that recovered the ideal-like Hall resistivity. 
Then, the correlation energy $E^{1/4}_{corr}=4.25456$ meV and the chemical potential $\mu_{1/4}=4.0182$ meV were obtained at $B_{1/4}=38.0$ T from 
$\mu_{1 / 4}=\xi(\frac{1}{2}-\zeta_{1/4})B_{1/4}-E^{1/4}_{corr}$.
The directly calculated chemical potential considering up to $\delta_3$ is pretty close to $\mu_{1 / 4}=4.140$ meV that yields the ideal-like Hall resistivity.
$\mu_{1/4}$ is taken as the initial value $\mu_\nu^0$ of the filling sequences $\nu = q/(4q-1)$, whereas the lower sequence $\nu = q/(4q+1)$ has $\mu_{1/6}$ as its $\mu_\nu^0$.  

On the other hand, the filling factors of $\nu=4/5, 5/7$, and unity included in Fig.~6(a) do not belong to the sequences $\nu = q/(2q\pm 1)$ which were studied in this study.
The former two are members of the sequences $\nu = (3q\pm1)/(4q\pm1)$, while the latter belongs to another class, namely the IQHE.
Although no data for energy gaps were available for these fillings, their Hall resistivities were added in Fig.~6(a) just for a continuation of the curve. 
Specifically, the filling factor $\nu=5/7$ belongs to the lower part of the sequences $\nu = (3q\pm1)/(4q\pm1)$, whose reference is $\nu=3/4$, while $\nu=4/5$ belongs to the upper part $\nu = (3q+1)/ (4q+1)$.
Thus, $\mu_{1/2}$ $(\mu_{3/4})$ is taken as the zero-field chemical potential for $\nu=5/7$ ($\nu=4/5$).   
The ideal-like Hall resistivity is obtained at filling $\nu=3/4$ in terms of $\mu_{3/4}=7.9863$ meV, $\zeta_{3/4}=0.2854$, and $\delta_n^{3/4}=0.431/[3\cdot 5\cdots(2n+1)]$.

Although the intercepts $\mu_{3/4}$ and $\mu_{1/2}$ were given, the slopes $\kappa_\nu$ for $\nu=4/5$ and $\nu=5/7$ are not easily obtained because of a lack of data on the energy gaps. 
For this reason, appropriate values for the parameters $\kappa_\nu$, $\zeta_\nu$, and $\delta_n^\nu$ for fillings $\nu=4/5$ and $\nu=5/7$ (Table I) were simply chosen in fitting the Hall resistivity curve.
Consequently, the parameter values for $\nu=5/7$ and $4/5$ given in Table I are less reliable than the values for other fillings.

The same situation happens in filling $\nu=1$, which is a member of the IQHE.
The state of filling unity has a different aspect from other fractional fillings of the LLL because the lowest excited state from $\nu=1$ belongs to the next Landau level ($N=0$, $\sigma=-1$), as shown in Fig.~5. Therefore, information on the energy gap and more careful treatment are needed for studying $\nu=1$. 
In this study, a higher chemical potential slope ($\kappa_1$) and larger $\mu_1^0$ were simply selected following the varying trend, and appropriate values of  $\delta_n^1$ were chosen to fit the Hall resistivity near unity filling. 

\begin{figure}[t]
\centering
\includegraphics[width=0.9\linewidth]{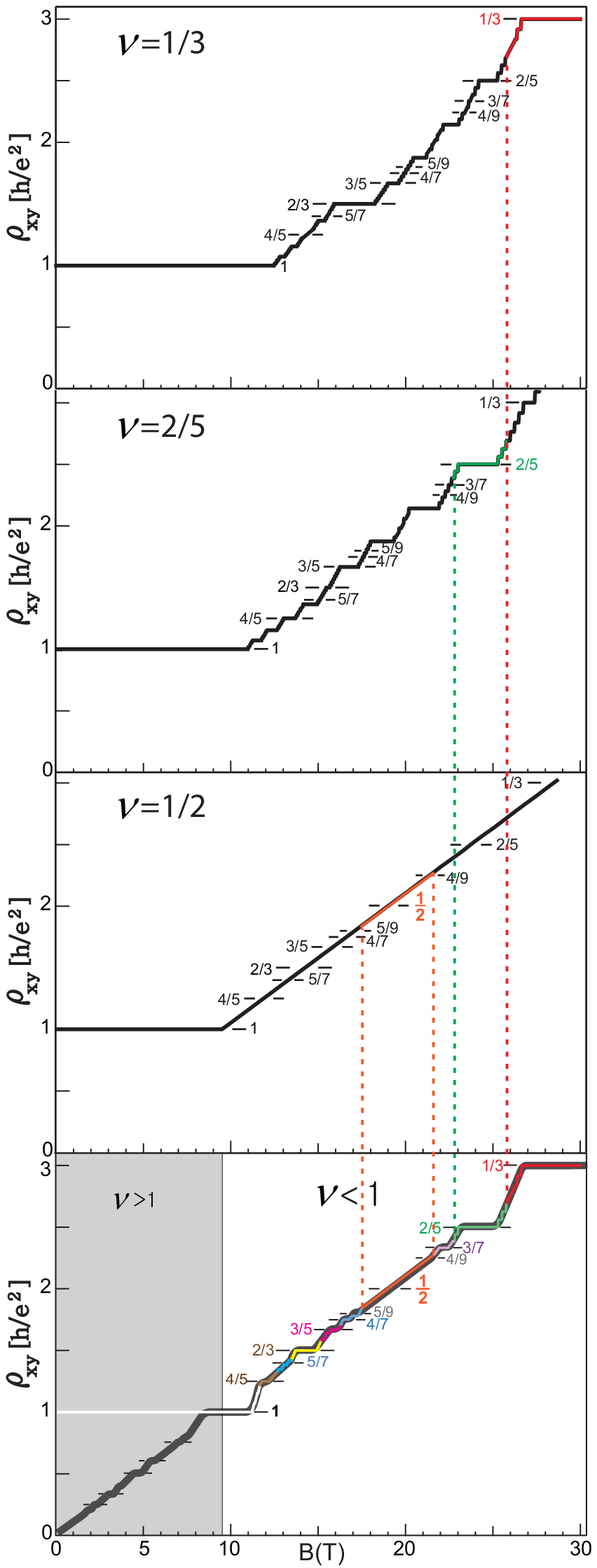}
\caption{The calculated Hall resistivity curves (black lines) were shown for fillings $\nu=1/3, 2/5$, and $1/2$ in the upper three panels. The measured Hall resistivity curve in the lowermost panel is reproduced by taking the corresponding part between two transition points. 
} 
\label{fig:Fig7}
\end{figure}

\section{Reproducing the Hall resistivity curve} 
The theoretical Hall resistivity curve was obtained separately for each plateau region using the parameter values given in Table I because each plateau state belongs to a different phase \cite{Sondhi}. The transition between 
adjacent plateau states is considered a quantum phase transition.

The procedure to obtain the Hall resistivity shown in Fig.~6(a) is as follows. 
The Hall resistivity curves obtained for the adjacent plateau states, such as two dark curves given in the two uppermost panels in Fig.~7, are superimposed, and the best point in the overlapped region is chosen as the transition point.
In this manner, the Hall resistivity curve $\rho_{xy}(B)$ over the entire domain of the magnetic field, $B = 9 - 30$ T, was obtained, as shown in the lowermost panel in Fig.~7, i.e., Fig.~6(a).



\end{document}